\documentclass[showpacs,preprintnumbers,amsmath,amssymb,preprint]{revtex4}

\usepackage{amsfonts}
\usepackage{amssymb}
\usepackage{graphicx}
\usepackage[cmtip,arrow]{xy}
\newlength{\vshift}
\newlength{\hshift}
\usepackage{amssymb,amsopn}

\begin{document}
\title{Weyl-Wigner-Moyal Formalism for Fermi Classical Systems\footnote{We devote this
paper to the memory of our friend Guillermo Moreno.}}

\author{ I. Galaviz}\email{igalaviz@fis.cinvestav.mx} \affiliation{Departamento de F\'{\i}sica,
Centro de Investigaci\'on y de Estudios Avanzados del IPN\\
P.O. Box 14-740, 07000 M\'exico D.F., M\'exico}

\author{H. Garc\'{\i}a-Compe\'an}
\email{compean@fis.cinvestav.mx} \affiliation{Departamento de
F\'{\i}sica,
Centro de Investigaci\'on y de Estudios Avanzados del IPN\\
P.O. Box 14-740, 07000 M\'exico D.F., M\'exico}
\affiliation{Centro de Investigaci\'on y de Estudios Avanzados del
IPN, Unidad Monterrey\\
Cerro de las Mitras 2565, cp. 64060, Col.
Obispado, Monterrey N.L., M\'exico}

\author{M. Przanowski}\email{przan@fis.cinvestav.mx}
\affiliation{Institute of Physics\\
Technical University of \L \'od\'z\\
W\'olcza\'nska 219, 93-005, \L \'od\'z, Poland}

\author{F.J. Turrubiates}\email{fturrub@fis.cinvestav.mx}
\affiliation{Departamento de F\'{\i}sica, Escuela Superior de F\'{\i}sica y Matem\'aticas del I.P.N.\\
Unidad Adolfo L\'opez Mateos, Edificio 9, 07738, M\'exico D.F.,
M\'exico}
\date{\today}

\begin{abstract}
The Weyl-Wigner-Moyal formalism of fermionic classical systems
with a finite number of degrees of freedom is considered. This
correspondence is studied by computing the relevant
Stratonovich-Weyl quantizer. The Moyal $\star$-product, Wigner
functions and normal ordering are obtained for generic fermionic
systems. Finally, this formalism is used to perform the
deformation quantization of the Fermi oscillator and the
supersymmetric quantum mechanics.
\end{abstract}
\vskip -1truecm
\maketitle

\vskip -1.3truecm
\newpage

\setcounter{equation}{0}

\section{Introduction}

The formalism introduced by H. Weyl \cite{weyl}, E.P. Wigner
\cite{wigner},  A. Groenewold \cite{groene} and J.E. Moyal
\cite{moyal} (or WWM formalism), establishes an isomorphism
between the Heisenberg operator algebra and the corresponding
algebra of symbols of these operators through the so called WWM
correspondence. The operator product is mapped to an associative
and noncommutative product called the Moyal $\star$-product.
Eventually, the theory of deformation quantization (from which the
WWM correspondence is an example) was introduced in 1978 by Bayen,
Flato, Fronsdal, Lichnerowicz and Sternheimer (BFFLS) \cite{bayen}
(for some recent reviews, see
\cite{zachosrev,reviewone,reviewtwo}). In BFFLS paper, deformation
quantization was introduced as an alternative procedure to the
canonical quantization and the path integral quantization in
quantum mechanics \cite{bayen}. Similarly to the path integral
formalism, deformation quantization uses the algebraic structure
of classical systems instead of operator theory.

This formalism has been employed also to quantize several physical
systems from particles to strings and recently used in the
quantization of dissipative systems \cite{paco}. The philosophy of
deformation quantization is based on the fact that quantization of
classical systems can be regarded equivalently as deformations of
the algebraic structures associated with these classical systems
\cite{stern}. Recently, WWM formalism has received a great deal of
interest motivated by the fact that it is contained inside string
theory. The presence of a non-zero constant $B$-field on the
worldvolume of a D-brane deforms the product of functions (or
classical fields) on the D-brane, and the ordinary effective gauge
theory on the brane turns out into a noncommutative field theory
with the usual Moyal $\star$-product \cite{sw}.

In deformation quantization, the quantization is understood as a
deformation of the algebra of classical observables and not as a
radical change of the mathematical nature of them. The ordinary
product of functions is then deformed into the Moyal
$\star$-product and the Poisson brackets turn into the Moyal
brackets. More generally by explicit construction $\star$-product
has been proved to exist for any finite dimensional symplectic
manifold \cite{prueba,fedosov}. One of the remarkable works is
that of Fedosov \cite{fedosov}, in where he was able to find a
general star product for any symplectic manifold with symplectic
geometry. Recently the geometric origin of the Fedosov's star
product, was elucidated in \cite{jaromir}). Then in 1997
Kontsevich \cite{k} proved the existence of the star product for
any Poisson manifold, and the explicit construction was done for
$\mathbb{R}^n$.

Deformation quantization formalism has a firm mathematical basis,
however its application to the quantization of arbitrary physical
systems presents still great challenges (see \cite{reviewtwo}).
Most of the cases studied by the deformation quantization for
systems with a finite number of degrees of freedom deal with
bosonic variables. However the analysis of some classical physical
systems requires the description of fermionic degrees of freedom
which involve Grassmann variables. These systems has been
discussed in the literature for some years
\cite{berezinbook,casa,marinov,sberezin,dewittbook,susyqmb}. The
canonical quantization of these systems by using fermionic quantum
mechanics was studied in \cite{marnelius}. Very recently some
authors started to applied the deformation quantization for these
classical fermionic systems
\cite{bordemann,bordemannII,duetsch,zachosfermions,hirshfield,clifford}.
In particular in Refs. \cite{hirshfield,clifford} the authors show
concretely how the deformation quantization program can be carried
over to specific physical systems. The same techniques have been
applied recently for the noncommutative superspace \cite{seiberg}.

In the present paper we study another approach to deformation
quantization for fermionic systems by employing the
Weyl-Wigner-Moyal correspondence \cite{wwmformalism,tata,hillery}.
We find that the super-Hilbert space structure gives some similar
formulas to the ones obtained within the WWM formalism for the
bosonic case. Moreover we will be able to obtain some of the
proposed basic formulas of the deformation quantization formalism
described in Ref. \cite{hirshfield} and generalize them. In this
respect, the present paper is complementary and provides some
support to Ref. \cite{hirshfield}.

Our paper is organized as follows. In sec. II we give general
prescriptions of quantum mechanical systems of fermions described
by $2n$ (odd) Grassmann variables. Sec. III is devoted to
construct the Stratonovich-Weyl (SW) quantizer which is the main
object to determine the Weyl correspondence between operators and
functions on the fermionic phase space. Then the SW quantizer for
Fermi systems is defined and its main properties are found. Using
a modified notion of the {\it trace} of an operator we show that
these properties have a similar form as the respective properties
in the bosonic case. The Moyal $\star$-product is found in sec. IV
and in sec. V the Wigner function for Fermi systems is also
obtained. Sec. VI is devoted to study the normal ordering for
generic fermionic systems. In sec. VII we discuss in detail two
examples: the Fermi oscillator and the supersymmetric harmonic
oscillator. Finally, in sec. VIII, some concluding remarks close
the paper.

\vskip 2truecm
\section{Preliminaries of Quantum Mechanics for Fermionic Systems}

In this section we review some important facts concerning
the quantization of fermionic classical systems with a finite
number of degrees of freedom. Our aim is not to provide an
extensive review but briefly recall some of their relevant
properties. We refer the reader to Weinberg's book \cite{wein}
for details.

Let $\psi=(\psi_1,\cdots,\psi_n)$ and $\pi=(\pi_1,\cdots,\pi_n)$
be complex Grassmann coordinates on the phase space
$\Gamma_F^{2n}$ of the relevant purely fermionic classical system.
The momenta $\pi_j$ are given by
\begin{equation}
\pi_j=i\psi_j^*, \ \ \ \ \ \ \  j=1,\cdots,n.
\end{equation}
(In what follows $j,k,l$ run from 1 to $n$.)

Quantization establishes the rules
\begin{equation}
\begin{array}{cc}
{[\widehat{\psi}_j,\widehat{\pi}_k]}_+ = i\hbar \delta_{jk}, \\
{[\widehat{\psi}_j,\widehat{\psi}_k]}_+ = 0 =
{[\widehat{\pi}_j,\widehat{\pi}_k]}_+,\\
{[\widehat{\psi}_j,\widehat{\psi}_k^*]}_+ = \hbar \delta_{jk},
\end{array}
\end{equation}
where $[\cdot,\cdot]_+$ stands for the anticommutator and
$[\cdot,\cdot]_-$ defines the commutator, i.e.
$[\widehat{A},\widehat{B}]_\pm:=\widehat{A}\cdot \widehat{B} \pm
\widehat{B} \cdot \widehat{A}$.

Define also
\begin{equation}
\widehat{b}_j:= {\widehat{\psi}_j \over \sqrt{\hbar}}, \ \ \ \ \ \
\ \ \ \  \ \ \ \ \ \ \widehat{b}_j^*:= {\widehat{\psi}_j^* \over
\sqrt{\hbar}}. \label{escalera}
\end{equation}
There exists a vacuum state given by the ket vector $|0\rangle$ or
a bra vector $\langle 0|$ defined by
\begin{equation}
\widehat{b}_j|0\rangle = 0, \ \ \ \ \ \  \ \ \ \ \ \ \  \langle 0|
\widehat{b}_j^* = 0, \ \ \ \ \ \ \ \ \ \  \forall j,
\end{equation}
satisfying the normalization condition $\langle
0|0 \rangle = 1.$ The basis of all states can be constructed from
excitations of the vacuum state $|0\rangle$ and they are given by
\begin{equation}
|j,k,l,\cdots \rangle :=\widehat{b}_j^* \widehat{b}_k
^*\widehat{b}_l^* \cdots |0\rangle .
\end{equation}
Then if $l\not\in\{j,k,\cdots\}$ one gets
\begin{equation}
\begin{array}{c}
\widehat{b}_l |j,k,\cdots\rangle = 0, \ \ \ \ \ \ \ \ \ \ \
\widehat{b}_l^* |j,k,\cdots\rangle = |l,j,k,\cdots\rangle.
\end{array}
\end{equation}
Moreover
\begin{equation}
\begin{array}{c}
\widehat{b}_l |l,j,k,\cdots\rangle = |j,k,\cdots\rangle,  \ \ \ \
\ \ \widehat{b}_l^* |l,j,k,\cdots\rangle = 0.
\end{array}
\end{equation}
The dual basis reads
\begin{equation}
\langle j,k,l,\cdots| := \langle 0 | \cdots \widehat{b}_l
\widehat{b}_k \widehat{b}_j.
\end{equation}
One quickly finds that
\begin{equation}
\langle {j_1},k_1,l_1,\cdots|j_2, k_2, l_2,\cdots \rangle =
\bigg\{%
\begin{array}{l}
  0 \mathrm{\ if\ } \{j_1,k_1,l_1,\cdots\}\neq \{j_2,k_2,l_2,\cdots\} \\
  1 \mathrm{\ if\ } j_1=j_2, k_1= k_2, l_1=l_2,\cdots
\end{array}%
\end{equation}
Now we look for the state $|\psi_1, \cdots, \psi_n \rangle \equiv
|\psi\rangle$ satisfying the following condition
\begin{equation}
\widehat{\psi}_j | \psi \rangle = \psi_j | \psi \rangle, \ \ \ \ \
\forall j. \label{eigenvalue}
\end{equation}
It is an easy matter to show that $| \psi \rangle$ has the
following form
\begin{equation}
|\psi\rangle = \exp\left\{-{i \over \hbar} \sum_{j=1}^n
\widehat{\pi}_j \psi_j\right\} |0 \rangle.
\label{seis}
\end{equation}
Indeed
(\ref{eigenvalue})
\begin{eqnarray}
\widehat{\psi}_k |\psi\rangle &=& \widehat{\psi}_k \exp \bigg\{
-{i \over \hbar} \widehat{\pi}_k\psi_k \bigg\}\cdot \exp \bigg\{
-{i
\over \hbar} \sum_{j\neq k} \widehat{\pi}_j\psi_j \bigg\} |0\rangle \nonumber\\
&=& \psi_k \exp \bigg\{ -{i \over \hbar} \sum_{j\neq k} \widehat{\pi}_j\psi_j \bigg\} |0\rangle \nonumber\\
&=& \psi_k | \psi\rangle. \label{vpropios}
\end{eqnarray}
In order to obtain the above result we have used the facts that
$\widehat{\psi}_k|0\rangle = 0$ and
$$
\widehat{\psi}_k \exp \bigg\{ -{i \over \hbar} \sum_{j\neq k}^n
\widehat{\pi}_j \psi_j \bigg\} = \exp \bigg\{ -{i \over \hbar}
\sum_{j\neq k} \widehat{\pi}_j \psi_j \bigg\} \widehat{\psi}_k,
$$
\begin{equation}
\psi_k\exp \bigg\{ -{i \over \hbar} \sum_{j\neq k} \widehat{\pi}_j
\psi_j \bigg\}= \psi_k\exp \bigg\{-{i\over \hbar} \sum_{j=1}^n
\widehat{\pi}_j \psi_j \bigg\}.
\end{equation}
Thus Eq. (\ref{seis}) holds true.

Then it is not difficult to get the crucial formula
\begin{equation}
\exp\bigg\{ -{i \over \hbar} \sum_{j=1}^n \widehat{\pi}_j
\xi_j\bigg\} |\psi\rangle = |\psi + \xi \rangle,\label{prop}
\end{equation}
which has the same form as the corresponging formula for bosonic
degrees of freedom \cite{wwmformalism,tata,hillery}, but in the
present case the order of $\widehat{\pi}_j \xi_j$ should be
respected.

Define
\begin{equation}
\begin{array}{l}
\langle \psi | := \langle 0| \widehat{\psi}_1 \cdots
\widehat{\psi}_n \exp\bigg\{ -{i \over \hbar} \sum_{j=1}^n \psi_j
\widehat{\pi}_j \bigg\}\\
= \langle 0| \big( \prod_{j=1}^n\widehat{\psi}_j\big) \exp\bigg\{
{i \over \hbar} \sum_{j=1}^n \widehat{\pi}_j \psi_j \bigg\}.
\end{array}
\label{dos}
\end{equation}
The ordering of $\widehat{\psi}_j$ can be arbitrary chosen but
then must be fixed. We choose the ordering $\widehat{\psi}_1\cdots
\widehat{\psi}_n$. One finds (compare with (\ref{vpropios})):
\begin{equation}
\langle \psi |\widehat{\psi}_j = \langle \psi |\psi_j,  \ \ \ \ \
\forall j.
\end{equation}
Comparing Eqs. (\ref{seis}) and (\ref{dos}) one can see that in
contrary to the bosonic case we have, $\langle \psi | \neq \big(
|\psi \rangle \big)^*.$ We note also the following important
point: The ket $|\psi\rangle$, by its very definition
(\ref{seis}), commutes with all Grassmann numbers $\eta$, i.e.,
$\eta|\psi\rangle = |\psi\rangle \eta.$ But for the bra $\langle
\psi |$ defined by (\ref{dos}) one has
\begin{equation}
\eta\langle\psi| = (-1)^{\varepsilon_{\eta}\cdot n}\langle \psi|
\eta,
\end{equation}
where $\varepsilon_{\eta}=1$ for odd Grassmann numbers or $\varepsilon_{\eta}=0$ for even
Grassmann numbers.
The inner product $\langle \psi' | \psi \rangle$, after some minor
computations, is found to read
\begin{equation}
\begin{array}{l}
\langle \psi'|\psi\rangle = \langle 0|\widehat{\psi}_1 \cdots
\widehat{\psi}_n \exp\bigg\{ {i \over \hbar} \sum_{j=1}^n
\widehat{\pi}_j\big(\psi_j'-\psi_j \big) \bigg\}|0\rangle\\
= \prod_{j=1}^n\big(\psi_j-\psi_j' \big)
=:\delta\big(\psi-\psi'\big). \label{tres}
\end{array}
\end{equation}
The analogous procedure can be performed for $|\pi\rangle$. We
look for $|\pi\rangle$ and $\langle\pi|$ such that
\begin{eqnarray}
\widehat{\pi}_j|\pi\rangle &=& \pi_j |\pi\rangle, \nonumber \\
\langle \pi| \widehat{\pi}_j &=& \langle \pi| \pi_j,   \ \ \ \ \ \forall j.
\end{eqnarray}

It is an easy matter to show that
\begin{equation}
|\pi\rangle = \exp\bigg\{ -{i\over \hbar} \sum_{j=1}^n
\widehat{\psi}_j\pi_j\bigg\}\prod_{j=1}^n\widehat{\pi}_j|0\rangle
\label{siete}
\end{equation}
and
\begin{equation}
\langle\pi| = \langle 0| \exp\bigg\{  -{i\over \hbar} \sum_{j=1}^n
\pi_j \widehat{\psi}_j \bigg\} = \langle 0| \exp \bigg\{{i\over
\hbar} \sum_{j=1}^n \widehat{\psi}_j\pi_j\bigg\}.
\label{cinco}
\end{equation}
Of course $\langle\pi|\not=(|\pi\rangle)^*.$
Then the inner product $\langle\pi'|\pi\rangle$ reads
\begin{equation}
\langle\pi'|\pi\rangle= \prod_{j=1}^n\big(\pi'_j-\pi_j \big) =
\delta\big(\pi'-\pi \big).
\label{cuatro}
\end{equation}
(Compare the formulas (\ref{tres}) and (\ref{cuatro})).

From Eqs. (\ref{seis}) and (\ref{cinco}) we get
\begin{equation}
\langle\pi|\psi\rangle= \exp\bigg\{- {i\over\hbar} \sum_{j=1}^n
\pi_j\psi_j \bigg\}.
\label{mixone}
\end{equation}
In a similar way using Eqs. (\ref{dos}) and (\ref{siete}) one obtains

\begin{equation}
\langle \psi|\pi \rangle = (i^n\hbar)^n \exp \bigg\{ {i\over\hbar}
\sum_{j=1}^n \pi_j \psi_j \bigg\}.
\end{equation}
In this paper we use the following convention for integrals
\begin{equation}
\int \psi_j d\psi_j= - \int d\psi_j \psi_j = 1, \ \ \ \
\int \pi_j d\pi_j= - \int d\pi_j \pi_j = 1, \ \ \ \   \forall j.
\label{ocho}
\end{equation}

This yields
\begin{equation}
\int|\psi\rangle {\cal D}\psi \langle \psi|=1, \label{nueve}
\end{equation}
where ${\cal{D}\psi}= {d\psi_n} \cdots {d\psi_1}.$

\noindent $\big[$Note the ordering. In Ref. \cite{wein}
$\cal{D}\psi$ would be denoted by ${\prod}_{j=1}^n d\psi_j$. For
instance in \cite{wein}, Eq. (\ref{nueve}) takes the form
$\int|\psi\rangle((-1)^n {\cal D}\psi) \langle \psi|=1$ but the
convention for integrals used in Weinberg's book is $\int d\psi_j'
\psi_j' = -\int\psi_j' d\psi_j'=1. \big]$

Analogously one proves that
\begin{equation}
\int |\pi\rangle {(-1)^n \cal{D} \pi }\langle\pi| = 1.
\label{unidad}
\end{equation}
After this brief survey we are ready to implement the WWM
formalism. Let's start with the Stratonovich-Weyl
quantizer.

\vskip 2truecm
\section{The Stratonovich-Weyl Quantizer}

Let $f=f(\pi,\psi)$ be a classical observable on the Grassmann
phase space $\Gamma_F^{2n}$. The Fourier transform of $f$ is
defined by
\begin{equation}
\widetilde{f}(\lambda,\mu) := \int f(\pi,\psi) \exp \bigg\{-{i}
\sum_{j=1}^n(
 \pi_j\lambda_j +  \psi_j \mu_j)\bigg\} \prod d\pi
 d\psi,
\label{sw1}
\end{equation}
where $\prod d\pi d\psi:= d\pi_1 d\psi_1\cdots d\pi_n d\psi_n$.\\

Then it is an easy matter to show that
\begin{eqnarray}
{f}(\pi,\psi) := \int \widetilde{f}(\lambda,\mu)
\exp\bigg\{{i}\sum_{j=1}^n(\pi_j\lambda_j +  \psi_j \mu_j)\bigg\}
\prod d\lambda d\mu.
\label{sw2}
\end{eqnarray}

The Fourier transform of $f(\pi,\psi)=1$ reads
\begin{eqnarray}
\widetilde{1}(\lambda,\mu)&=& \int \exp \bigg\{-i
\sum_{j=1}^n(\pi_j \lambda_j + \psi_j
\mu_j)\bigg\} \prod d\pi d\psi \nonumber \\
&=&\int \mu_1\lambda_1 \psi_1\pi_1\cdots
\mu_n\lambda_n\psi_n\pi_n \prod d\pi d\psi \nonumber \\
&=& \mu_1\lambda_1 \cdots \mu_n\lambda_n= \delta(\mu,\lambda),
\label{sw6}
\end{eqnarray}
where $ \delta(\mu,\lambda)
:=(-1)^{{n(n-1)}\over{2}}\delta(\mu)\delta(\lambda).$ Thus we have
\begin{equation}
f(\lambda,\mu)= \int f(\lambda',\mu')
\delta(\mu-\mu',\lambda-\lambda') \prod d\lambda' d\mu'.
\label{sw7}
\end{equation}
Now we are at the position to consider the Stratonovich-Weyl
quantizer.

Let $f=f(\pi,\psi)$ be a smooth function on $\Gamma_F^{2n}$. Then
the Weyl quantization rule is given by
$$
\widehat{f} := \int \widetilde{f}(\lambda,\mu)
\exp\bigg\{i\sum_{j=1}^n(\widehat{\pi}_j\lambda_j +
\widehat{\psi}_j \mu_j)\bigg\} \prod d\lambda d\mu,
$$
where $\widetilde{f}(\lambda,\mu)$ is defined by Eq. (\ref{sw1}).
In another form
\begin{equation}
\widehat{f} = \int f(\pi,\psi) \widehat{\Omega}(\pi,\psi) \prod
d\pi d\psi,
\label{sw9}
\end{equation}
where
\begin{equation}
\widehat{\Omega}(\pi,\psi) =  \int \exp\bigg\{i\sum_{j=1}^n
\bigg[(\widehat{\pi}_j-\pi_j) \lambda_j +
(\widehat{\psi}_j-\psi_j)\mu_j  \bigg]\bigg\} \prod d\lambda d\mu,
\label{sw10}
\end{equation}
is called the {\it Stratonovich-Weyl (SW) quantizer} (compare with
Ref. \cite{sberezin}).

The SW quantizer can be rewritten in the following form
$$
\widehat{\Omega}(\pi,\psi) = \int
\exp\bigg\{-i\sum_{j=1}^n(\pi_j\lambda_j + \psi_j\mu_j)\bigg\}
\exp\bigg\{i\sum_{j=1}^n\widehat{\pi}_j\lambda_j\bigg\}
$$
$$
\times \exp\bigg\{i\sum_{j=1}^n\widehat{\psi}_j\mu_j\bigg\}
\exp\bigg\{ -{i\hbar\over 2} \sum_{j=1}^n\lambda_j\mu_j \bigg\}
\prod d\lambda d\mu
$$
\begin{equation}
 = \int \exp\bigg\{-i\sum_{j=1}^n(\pi_j\lambda_j +\psi_j\mu_j)\bigg\}
\exp\bigg\{i\sum_{j=1}^n\widehat{\psi}_j\mu_j\bigg\}
\exp\bigg\{i\sum_{j=1}^n\widehat{\pi}_j\lambda_j\bigg\} \exp
\bigg\{ {i\hbar\over 2}\sum_{j=1}^n\lambda_j\mu_j \bigg\} \prod
d\lambda d\mu.
\label{sw11}
\end{equation}

Using Eqs. (\ref{prop}) and (\ref{nueve}) we reexpress the above
equation in the form
$$
\widehat{\Omega}(\pi,\psi) =\int \exp\bigg\{-i\sum_{j=1}^n\pi_j
\lambda_j \bigg\} i^n \bigg[\psi'_1 -(\psi_1 + {\hbar\lambda_1
\over 2}) \bigg]\cdots \bigg[\psi'_n -(\psi_n + {\hbar\lambda_n
\over 2}) \bigg]
$$
$$
 \times (-1)^{n(n-1)\over 2} \cdot (-1)^{n(n+1)\over 2} i^n
\mu_1\cdots \mu_n {d\mu_1\cdots d\mu_n d\lambda_n \cdots
d\lambda_1} |\psi' - \hbar \lambda \rangle {\cal{D}\psi'} \langle
\psi' |
$$
$$
=\int{\cal{D}\lambda} \exp\bigg\{-i\sum_{j=1}^n\pi_j \lambda_j \bigg\}
\bigg[(\psi_1 + {\hbar\lambda_1 \over 2}) -\psi'_1 \bigg]\cdots \bigg[(\psi_n +
{\hbar\lambda_n \over 2}) - \psi'_n \bigg] {d\psi'_n \cdots
d\psi'_1} (-i)^n |\psi' - \hbar \lambda \rangle  \langle \psi' |
$$
\begin{equation}
= i^n\int{\cal{D}\lambda} \exp\bigg\{-i\sum_{j=1}^n\pi_j\lambda_j
\bigg\}|\psi - {\hbar \lambda\over 2} \rangle  \langle \psi +
{\hbar \lambda\over 2}|, \label{sw12}
\end{equation}
where ${\cal D} \lambda := d\lambda_n \dots d \lambda_1$. Thus we
arrive at a result very similar to that of bosonic case.

Analogously we have from Eqs. (\ref{siete}), (\ref{unidad}) and
(\ref{sw11}) that
$$
\widehat{\Omega}(\pi,\psi) =  (-1)^n\int \widehat{\Omega}
(\pi,\psi)|\pi' \rangle  {\cal{D}\pi'} \langle \pi'|
$$
\begin{equation}
 =(-i)^n \int
{\cal D}\mu \exp\bigg\{-i\sum_{j=1}^n\psi_j\mu_j\bigg\} \big|\pi-
{\hbar \mu\over 2} \big\rangle \big\langle \pi + {\hbar \mu\over
2}\big|, \label{sw13}
\end{equation}
where ${\cal D} \mu := d\mu_n \dots d \mu_1$. The formulas
(\ref{sw12}) and (\ref{sw13}) are similar to the corresponding
formulas for the bosonic case \cite{wwmformalism,tata,hillery}.

Define the following mapping `${\rm tr}$' which will be
called the ``trace" as
\begin{equation}
{\rm tr} \{\widehat{A} \} := c\int {\cal D}\psi \langle \psi |
\widehat{A} | \psi \rangle, \label{traza}
\end{equation}
where $c$ is to be determined from the condition
\begin{equation}
{\rm tr} \{\widehat{\Omega}(\pi,\psi)\} = 1. \label{cond}
\end{equation}
By using Eqs. (\ref{sw12}), (\ref{traza}) and (\ref{cond}) one
finds that $c=(i\hbar)^{-n}.$ Finally, the ``trace" of an operator
$\widehat{A}$ is
\begin{equation}
{\rm tr}\big\{ \widehat{A}\big\} = (i\hbar)^{-n} \int {\cal D}\psi
\langle \psi |\widehat{A} | \psi \rangle. \label{trace}
\end{equation}
In terms of the $\langle \pi |$-representation one has
\begin{equation}
{\rm tr}\big\{ \widehat{A}\big\} = (i\hbar)^{-n} \int {\cal D}\pi
\langle \pi |\widehat{A} | \pi \rangle. \label{traceone}
\end{equation}
With the above definition, Eq. (\ref{cond}) is also satisfied. In
addition, with the aid of Eq. (\ref{sw12}) and after laborious
computations we find the following important property of the
Stratonovich-Weyl quantizer
$$
{\rm tr}\bigg\{ \widehat{\Omega}(\pi',\psi')
\widehat{\Omega}(\pi'',\psi'') \bigg\} = (\psi'_1 - \psi''_1)(
\pi'_1- \pi''_1) \cdots (\psi'_n - \psi''_n)( \pi'_n- \pi''_n)
$$
\begin{equation}
= \delta(\psi'- \psi'', \pi' - \pi''). \label{ddtraza}
\end{equation}
This is a crucial formula which leads to the realization of the
Weyl correspondence in the case of classical systems of
fermions. We consider the Weyl correspondence
$\widehat{f}=W^{-1}(f)$ as given by
\begin{equation}
\widehat{f} = \int f(\pi',\psi') \widehat{\Omega}(\pi',\psi')
\prod d\pi' d\psi'.
\label{oper}
\end{equation}
Then, multiplying this equation by $\widehat{\Omega}(\pi,\psi)$,
taking its ``trace" and using Eq. (\ref{ddtraza}) we obtain
$$
{\rm tr} \bigg\{ \widehat{f}\widehat{\Omega}(\pi,\psi) \bigg\} =
\int {\rm tr} \bigg\{ \widehat{\Omega}(\pi',\psi')
\widehat{\Omega}(\pi,\psi) \bigg\} \prod d\pi' d\psi'
f(\pi',\psi')
$$
$$
= \int \delta(\psi'- \psi, \pi' - \pi) \prod d\pi' d\psi'
f(\pi',\psi') = f(\pi,\psi).
$$
Finally
\begin{equation}
f(\pi,\psi) = {\rm tr} \bigg\{
\widehat{f}\widehat{\Omega}(\pi,\psi) \bigg\}
\end{equation}
and this is exactly the same expression as in the bosonic case.

\noindent $\big[$ Remark: Similarly to the bosonic case one
introduces the oscillator variables $Q_j$ and $P_j$
\begin{eqnarray}
Q_j:={{1}\over{\sqrt{2}}}(\psi_j - i\pi_j)={{1}\over{\sqrt{2}}}(\psi_j+\psi_j^*),  \nonumber \\
P_j:={{1}\over{\sqrt{2}}}(\pi_j -
i\psi_j)={{i}\over{\sqrt{2}}}(\psi_j^*-\psi_j), \label{sw3}
\end{eqnarray}
which form $2n$ real coordinates of $\Gamma_F^{2n}$. From Eqs.
(\ref{ocho}) and (\ref{sw3}) we find the integrals
\begin{equation}
\begin{array}{c}
\int Q_jdQ_j = -\int dQ_j Q_j =1, \ \ \ \ \int P_jdP_j =-\int dP_j P_j =1, \ \ \ \ \forall j, \\
\prod d\pi d\psi = \prod dPdQ. \label{sw4}
\end{array}
\end{equation}

One can rewrite all the formulas in terms of $Q,P$. In particular
the Fourier integral is given by
\begin{equation}
\widetilde{f}(\alpha,\beta) = \int f(P,Q)
\exp\bigg\{-i\sum_{j=1}^n(P_j\alpha_j +  Q_j \beta_j)\bigg\} \prod
dP dQ,
\end{equation}
where
\begin{equation}
f(P,Q)=\int \widetilde{f}(\alpha,\beta)
\exp\bigg\{i\sum_{j=1}^n(P_j\alpha_j +  Q_j \beta_j)\bigg\} \prod
d\alpha d\beta. \label{sw5}
\end{equation}
While the Weyl correspondence in these variables reads
\begin{equation}
\widehat{f} = \int
f(\pi(P,Q),\psi(P,Q))\widehat{\Omega}(\pi(P,Q),\psi(P,Q)) \prod
dPdQ \label{sww}
\end{equation}
and
\begin{equation}
f(P,Q)= {\rm tr} \bigg\{ \widehat{f} \widehat{\Omega}(P,Q)
\bigg\}.
\end{equation}
Some of the above formulas will be used in
Sec. VI. $\big]$

Consider now ${\rm tr}\{\widehat{\Omega} (\pi',\psi') \widehat{\Omega} (\pi'',\psi'')
\widehat{\Omega}(\pi,\psi) \}$. As will be seen in the next section this is
the main point to find the Moyal product.
First
$$
\widehat{\Omega} (\pi',\psi') \widehat{\Omega} (\pi'',\psi'')
\widehat{\Omega}(\pi,\psi)  = \big({i\hbar\over 2}\big)^{3n} \int
\exp \bigg\{ -{2i \over \hbar} \big(\pi' \lambda' + \pi''
\lambda'' + \pi \lambda \big) \bigg\} |
\psi' - \lambda' \rangle  \langle \psi' + \lambda'  | \psi'' - \lambda'' \rangle  \\
$$
$$
 \times\langle \psi'' + \lambda'' | \psi - \lambda \rangle
\langle \psi + \lambda  | {\cal D}\lambda'  {\cal D}\lambda''
{\cal D}\lambda.
$$
Then
\begin{equation}
\begin{array}{l}
{\rm tr} \bigg\{ \widehat\Omega (\pi',\psi') \widehat{\Omega}
(\pi'',\psi'') \widehat{\Omega}(\pi,\psi) \bigg\} = (i\hbar)^{-n}
\big({i\hbar\over 2}\big)^{3n} \int {\cal D} \psi'''  \exp \bigg\{
-{2i \over \hbar} \big(\pi' \lambda' + \pi'' \lambda'' + \pi
\lambda \big) \bigg\} \langle \psi'''| \psi' - \lambda'
\rangle\\
\hspace{1.8cm} \times \langle \psi' + \lambda'  | \psi'' -
\lambda'' \rangle \langle \psi'' + \lambda'' | \psi - \lambda
\rangle (-1)^n\delta [\lambda- (\psi'''-\psi)] {\cal D}\lambda
{\cal D}\lambda' {\cal D}\lambda''  \\
= (-1)^n(i\hbar)^{-n}  \big({i\hbar\over 2}\big)^{3n} \int {\cal
D} \psi'''  \exp \bigg\{ -{2i \over \hbar} \big(\pi' \lambda' +
\pi'' \lambda'' +
\pi (\psi'''-\psi) \big) \bigg\} \langle \psi'''| \psi' - \lambda' \rangle\\
\hspace{1.8cm} \times \langle \psi' + \lambda'  | \psi'' -
\lambda'' \rangle (-1)^n \delta[ \lambda'' - ( 2\psi - \psi''' -
\psi'')] (-1)^n  {\cal D}\lambda''{\cal D}\lambda'   \\
= (-1)^n(i\hbar)^{-n}  \big({i\hbar\over 2}\big)^{3n} \int {\cal
D} \psi'''  \exp \bigg\{ -{2i \over \hbar} \big(\pi' \lambda' +
\pi'' ( 2\psi - \psi''' -
\psi'') + \pi (\psi'''-\psi) \big) \bigg\} \\
\hspace{1.8cm} \times \langle \psi'''| \psi' - \lambda' \rangle
(-1)^n \delta[ \lambda' - (2\psi'' -  2\psi + \psi''' - \psi' ) ] {\cal D}\lambda'   \\
=  \big({i\hbar\over 2}\big)^{2n} \exp \bigg\{ -{2i \over \hbar}
\big(\pi' (\psi'' -  \psi ) + \pi'' ( \psi - \psi' ) + \pi ( \psi'
- \psi'' ) \big) \bigg\}. \label{tresomegas}
\end{array}
\end{equation}

Employing (\ref{nueve}) and (\ref{trace}) one gets
$$
{\rm tr}\bigg\{ \widehat{A}\widehat{B} \bigg\} = (i\hbar)^{-n}
\int {\cal D} \psi \langle \psi | \widehat{A}\widehat{B} | \psi
\rangle
$$
$$
= (i\hbar)^{-n} \int {\cal D} \psi \langle \psi | \widehat{A} |
\psi'\rangle {\cal D} \psi' \langle \psi' | \widehat{B} | \psi
\rangle
$$
$$
= (i\hbar)^{-n} (-1)^{{\varepsilon_{\widehat{A}}}
{\varepsilon_{\widehat{B}}} + n {\varepsilon_{\widehat{B}}}}
(-1)^{n{\varepsilon_{\widehat{B}}}} \int {\cal D} \psi' \langle
\psi' | \widehat{B} | \psi \rangle  {\cal D} \psi \langle \psi |
\widehat{A} | \psi'\rangle
$$
$$
= (-1)^{{\varepsilon_{\widehat{A}}} {\varepsilon_{\widehat{B}}}}
{\rm tr} \bigg\{ \widehat{B} \widehat{A} \bigg\}.
$$
Thus we have the following relation
\begin{equation}
{\rm tr}\bigg\{ \widehat{A}\widehat{B} \bigg\} =
(-1)^{{\varepsilon_{\widehat{A}}} {\varepsilon_{\widehat{B}}}}
{\rm tr} \bigg\{ \widehat{B} \widehat{A} \bigg\},
\end{equation}
where $\varepsilon_{\widehat{A}}$, $\varepsilon_{\widehat{B}}$ are
the Grassmann parity of $\widehat{A}$ and $\widehat{B}$,
respectively. $\varepsilon_{\widehat{A}}=0,1$,
$\varepsilon_{\widehat{B}}=0,1$.
By the definition (\ref{sw10}), SW quantizer
is an even operator, i.e., $\varepsilon_{\widehat{\Omega}}=0$,
thus one gets
\begin{equation}
f(\pi,\psi) = {\rm tr} \bigg\{
\widehat{f}\widehat{\Omega}(\pi,\psi) \bigg\} = {\rm tr}\bigg\{
\widehat{\Omega}(\pi,\psi) \widehat{f} \bigg\}. \label{fun}
\end{equation}
\vskip 2truecm
\section{The Moyal Product}

In this section we find the Moyal $\star$-product. Let
$f=f(\pi,\psi)$ and $g=g(\pi,\psi)$ be any pair of functions
defined on our fermionic phase space $\Gamma_F^{2n}$ and let
$f=W^{-1}(\widehat{f})$ and $g=W^{-1}(\widehat{g})$ be the
corresponding operators via the Weyl correspondence. Then we are
looking for the product $(f\star g)(\pi,\psi)$ which corresponds
to the product of operators $\widehat{f} \cdot \widehat{g}$
through the Weyl correspondence

\begin{equation}
(f\star g)(\pi,\psi) = {\rm tr} \bigg\{ \widehat{f} \widehat{g}
\widehat{\Omega}(\pi,\psi) \bigg\}.
\label{star}
\end{equation}
Substituting (\ref{sw9}) into (\ref{star}) and using
(\ref{tresomegas}), we get
$$
(f\star g) (\pi,\psi) = \big({i\hbar\over 2}\big)^{2n} \int
f(\pi',\psi') g(\pi'',\psi'')  \exp \bigg\{ -{2i \over \hbar}
\bigg[i' (\psi'' - \psi ) + \pi'' ( \psi - \psi' ) + \pi ( \psi' -
\psi'' ) \bigg] \bigg\} {\cal D} \pi' {\cal D}\psi' {\cal D} \pi''
{\cal D}\psi''.
$$
By changing the variables: $\Psi' = \psi' - \psi$, $\Pi' = \pi' -
\pi$,  $\Psi'' = \psi'' -  \psi$, $ \Pi'' = \pi'' - \pi$, then the
Moyal product take the form
\begin{equation}
(f\star g)(\pi,\psi)= \big({i\hbar\over 2}\big)^{2n}\int
f(\Pi'+\pi,\Psi' + \psi)   g(\Pi'' + \pi,\Psi'' + \psi)  \exp
\bigg\{ -{2i \over \hbar} \big[ \Pi'\Psi'' - \Pi''\Psi' \big]
\bigg\} {\cal D} \Pi' {\cal D}\Psi' {\cal D} \Pi'' {\cal D}\Psi''.
\label{moyalcomp}
\end{equation}
Expanding $f$ and $g$ into the Taylor series and performing some manipulations one arrives
to the main formula
\begin{equation}
(f\star g)(\pi,\psi) = f(\pi,\psi) \exp \bigg\{ {i\hbar\over 2}
\buildrel{\leftrightarrow}\over{\cal P}_F\bigg\} g(\pi,\psi),
\label{moyalproduct}
\end{equation}
where
\begin{equation}
\buildrel{\leftrightarrow}\over{\cal P}_F=
{\overleftarrow{\partial } \over \partial\pi}
{\overrightarrow{\partial} \over \partial \psi} +
{\overleftarrow{\partial} \over \partial \psi}
{\overrightarrow{\partial} \over
\partial\pi}.
\end{equation}
with $\stackrel{\leftarrow}{\partial}$ and
$\stackrel{\rightarrow}{\partial}$ stand for the right derivative
and left derivative, respectively. This is the Moyal
$\star$-product for the fermionic part of the super-Poisson
bracket discussed in Refs.
\cite{casa,marinov,sberezin,dewittbook,hirshfield,bordemann,susyqmb,bordemannII}.

\vskip 2truecm
\section{The Wigner Function}

From Eq (\ref{fun}) one can conclude that the Wigner function in
the fermionic case should be defined simply by
\begin{equation}
\rho_W(\pi,\psi) = {\rm tr}\bigg\{
\widehat{\rho}\widehat{\Omega}(\pi, \psi) \bigg\},
\label{wignerone}
\end{equation}
where $\widehat{\rho}$ stands for the density operator.
Observe that the formula (\ref{wignerone}) is
almost the same as in the bosonic case \cite{wwmformalism}. In the
fermionic case we don't have a factor $1 \over (2\pi\hbar)^n$.
However, this factor in a modified form appears in the definition
of trace "${\rm tr}$" by Eqs. (\ref{trace}) or (\ref{traceone}).
Substituting Eqs. (\ref{sw12}) and (\ref{trace}) into
(\ref{wignerone}), after some computations we get
$$
\rho_W (\pi,\psi) = (i\hbar)^{-n} \int {\cal D}\psi' \langle \psi'
| \widehat{\rho}\widehat{\Omega}(\pi,\psi) | \psi' \rangle
$$
$$
=\hbar^{-n} \int {\cal D}\psi' {\cal D}\lambda \exp \bigg\{-i
\sum_{j=1}^n\pi_j\lambda_j \bigg\} (-1)^{n+ n \cdot
\varepsilon_{\widehat{\rho}}} \langle \psi' | \widehat{\rho}  |
\psi - {\hbar \lambda \over 2}\rangle \delta(\psi' - (\psi +
{\hbar \lambda \over 2} ))
$$
$$
= \hbar^{-n} \int {\cal D}\lambda \exp \bigg\{-i \sum_{j=1}^n
\pi_j \lambda_j \bigg\} \langle \psi + {\hbar \lambda \over 2} |
\widehat{\rho} | \psi - {\hbar \lambda \over 2}\rangle.
$$
Thus, finally
\begin{equation}
\rho_W (\pi,\psi) = \hbar^{-n} \int {\cal D}\lambda \exp \bigg\{-i
\sum_{j=1}^n \pi_j \lambda_j \bigg\}  \langle \psi + {\hbar
\lambda \over 2} | \widehat{\rho} | \psi - {\hbar \lambda \over
2}\rangle.
\end{equation}

\vskip 2truecm
\section{Normal Ordering}

Assume first $n=1$ and consider $f:=b^*b$ and $g:=bb^*$, where $b,b^*$ are defined by
$b={\psi \over \sqrt{\hbar}}$ and $b^*={\psi^* \over
\sqrt{\hbar}}$. Then the quantization of $f$ and $g$ in the {\it
Berezin-Wick (B-W)} or {\it normal ordering} gives
\begin{eqnarray}
f=b^*b {\buildrel {B-W} \over \mapsto} \widehat{b}^* \widehat{b},
\ \ \ \ \ \ \ \ \ \ \  g=bb^* {\buildrel {B-W} \over \mapsto} -
\widehat{b}^* \widehat{b}.
\end{eqnarray}

It is an easy matter to find that the same results are obtained if
one uses the Weyl rule (\ref{sw9}) or (\ref{sww})
changing $f$ and $g$ by introducing the operator of normal ordering
$\widehat{\cal N}$. Thus
\begin{eqnarray}
\widehat{b}^* \widehat{b}&=& \int (\widehat{\mathcal{N}} b^*b)
\widehat{\Omega}(\pi,\psi) d\pi d\psi, \\ -\widehat{b}^*
\widehat{b}&=& \int (\widehat{\mathcal{N}} bb^*)
\widehat{\Omega}(\pi,\psi) d\pi d\psi,
\end{eqnarray}
where
\begin{equation}
 \widehat{\mathcal{N}}:=
\exp\left\{ {i\hbar \over 2}{\overrightarrow{\partial}^2 \over
\partial\psi
\partial \pi}\right\}.
\label{normalo}
\end{equation}
By induction one arrives at the general result for any $n$. Let
$f(\pi,\psi)$ be any function on $\Gamma_F^{2n}$. Then the {\it
Berezin-Wick quantization} ({\it normal quantization}) of $f$ is
done by the Weyl quantization of a modified $f$ according to
\begin{equation}
f_{\mathcal{N}}:= \widehat{\mathcal{N}}f,
\end{equation}
where
\begin{eqnarray}
\widehat{\mathcal{N}}&=& \exp\left\{ {i\hbar \over 2}
\sum_{j=1}^{n} {\overrightarrow{\partial}^2 \over \partial\psi_{j}
\partial \pi_{j}}\right\} = \exp\left\{ {\hbar \over 2}
\sum_{j=1}^{n}{\overrightarrow{\partial}^2 \over \partial\psi_{j}
\partial \psi^*_{j}}\right\}\nonumber\\
&=&\exp\left\{ {1 \over 2}\sum_{j=1}^{n}{\overrightarrow{\partial}^2 \over
\partial b_{j}\partial b^*_{j}}\right\} = \exp\left\{ {i\hbar \over 2}
\sum_{j=1}^{n}{\overrightarrow{\partial}^2 \over \partial Q_{j}
\partial P_{j}}\right\}.
\label{noordering}
\end{eqnarray}
Then the Weyl correspondence reads
\begin{equation}
\widehat{f}_{\mathcal{N}}=\int (\widehat{\mathcal{N}}f)
\widehat{\Omega} \prod d\pi d\psi.
\end{equation}
In the next section we will analyze two simple examples to see how
the WWM formalism works for systems involving fermions.

\vskip 2truecm
\section{Examples}

\subsection{Fermi Oscillator}

As an example of our WWM formalism consider the fermionic
oscillator for one degree of freedom ($n=1$). The Lagrangian
of such an oscillator reads
\begin{equation}
L = i \psi^* \dot{\psi} - \omega \psi^* \psi.
\end{equation}
Then the momentum conjugate to $\psi$ is $\widetilde{\pi} =
{{\buildrel \rightarrow \over \partial}L \over
\partial \dot{\psi} } = -i \psi^*,$ while the Hamiltonian $H$ is given by
\begin{equation}
H = \dot{\psi} \widetilde{\pi} - L = i\omega \widetilde{\pi}\psi.
\label{hamiltonian}
\end{equation}
Here $\stackrel{\to}{\partial}$ stands for the left-derivative
\cite{berezinbook,hirshfield}.

Now is convenient to use $\pi:= -\widetilde{\pi}$ and $\pi = i
\psi^*$ instead of $\widetilde{\pi}$. Thus in terms of $\pi$ the
Hamiltonian (\ref{hamiltonian}) reads
\begin{equation}
H= -i\omega \pi\psi = \omega \psi^*\psi.
\label{ham_omega}
\end{equation}

The coherent state $|\psi\rangle$ (\ref{seis}) is given by
\begin{equation}
|\psi\rangle = \exp\left \{ -{i \over \hbar} \widehat{\pi}
\psi\right \} | 0 \rangle =  \exp\left \{{\widehat{\psi}^* \psi
\over \hbar} \right\} | 0 \rangle, \label{coherent state}
\end{equation}
where $|0 \rangle$ is the ground state and $\widehat{\pi}$ is the
momentum operator. The quantum operators satisfy the well known
anti-commutation relations ${[\widehat{\psi},\widehat{\pi}]}_+ =
i\hbar,$ or ${[\widehat{\psi},\widehat{\psi}^*]}_+ = \hbar.$

In the matrix representation
\begin{equation}
\begin{array}{ccc}
|0 \rangle = \left(%
\begin{array}{cc}
  1 \\
  0 \\
\end{array}%
\right), & \ \ \ \ \ \ \ \widehat{\psi}=\sqrt{\hbar} \left(%
\begin{array}{cc}
  0 & 1 \\
  0 & 0 \\
\end{array}%
\right), & \ \ \ \ \ \ \ \widehat{\psi}^*=\sqrt{\hbar} \left(%
\begin{array}{cc}
  0 & 0 \\
  1 & 0 \\
\end{array}%
\right).
\end{array}
\label{represent}
\end{equation}
Then the vector $|1\rangle$  is defined by $|1\rangle = {1\over
\sqrt{\hbar}} \widehat{\psi}^*|0\rangle$ or in the matrix
representation $|1\rangle = \left(%
\begin{array}{c}
  0 \\
  1 \\
\end{array}%
\right)$. In fact it is a supervector with Grassmann parity
$\varepsilon_{_{|1\rangle}}=1$ or an ``odd supervector''.
Obviously the state $|0\rangle$ has parity
$\varepsilon_{_{|0\rangle}}=0$. In other words it is an ``even
supervector''. The state $|\psi\rangle$ has also the parity
$\varepsilon_{_{| \psi \rangle}}=0$ (it is an even supervector).
By (\ref{coherent state}) and (\ref{represent}) we have
\begin{equation}
|\psi \rangle = \exp\left\{ {\widehat{\psi}^* \psi \over \hbar}
\right\} | 0 \rangle =\left( 1-{\psi \widehat{\psi}^* \over \hbar}
\right) |0\rangle = \left(%
\begin{array}{c}
  1 \\
  -{\psi \over \sqrt{\hbar}} \\
\end{array}%
\right) = |0\rangle - {\psi \over \sqrt{\hbar}} |1\rangle.
\label{state}
\end{equation}
The dual vector is given by $ \langle \psi | = \sqrt{\hbar} (0,1)
- \psi (1,0) = \sqrt{\hbar} \langle 1| - \psi \langle 0|.$
Remember that $\langle \psi| \psi' \rangle = \psi' - \psi = \delta
(\psi' - \psi)$.  Then the complementary states are
\begin{equation}
\langle \pi | = \langle \psi^* | := (|\psi\rangle)^* = (1,\
-{\psi^* \over \sqrt{\hbar}}) = \langle 0 | - {\psi^* \over
\sqrt{\hbar}} \langle 1 |.
\end{equation}
Of course $\langle 0 | = (|0\rangle)^* = (1,\ 0),$ and $\langle 1|
= (|1\rangle)^* = (0,\ 1).$ We have also from Eq. (\ref{siete})
(for $n=1$) that
\begin{equation}
|\pi\rangle = \exp\left\{ -{i\over \hbar} \widehat{\psi}\pi
\right\}\widehat{\pi} |0 \rangle = i \sqrt{\hbar}|1\rangle - i
\psi^* |0
\rangle =  i \sqrt{\hbar} \left(%
\begin{array}{c}
  0 \\
  1 \\
\end{array}%
\right) - i \psi^* \left(%
\begin{array}{c}
  1 \\
  0 \\
\end{array}%
\right).
\end{equation}
It is also an easy matter to see that the state $|\pi\rangle$ has odd
parity and therefore $\varepsilon_{|\pi\rangle} = 1$. Simple calculations show that
$\int | \psi \rangle d\psi\langle \psi | =1$ and $\int | \pi \rangle (-1)d\pi \langle \pi |
=1$, as expected.

\vskip 1 truecm
\noindent
{\it Stratonovich-Weyl Quantizer}

According to our previous considerations the Stratonovich-Weyl
quantizer is given by Eq. (\ref{sw12})
$$
\widehat{\Omega} (\pi,\psi) = i \int d\lambda \exp\{ -i\pi\lambda
\} |\psi -{\hbar\lambda \over 2} \rangle \langle \psi +
{\hbar\lambda \over 2} |
$$
$$
=i \int d\lambda (1+\psi^*\lambda) \bigg[ \left(%
\begin{array}{c}
  1 \\
  0 \\
\end{array}%
\right) - {(\psi - {\hbar\lambda\over 2}) \over \sqrt{\hbar}} \left(%
\begin{array}{c}
  0 \\
  1 \\
\end{array}%
\right)\bigg] \otimes \bigg[\sqrt{\hbar} \left(%
\begin{array}{cc}
  0 & 1 \\
\end{array}%
\right) - {(\psi + {\hbar\lambda\over 2}) } \left(%
\begin{array}{cc}
  1 & 0 \\
\end{array}%
\right)\bigg].
$$
Integrating over $d\lambda$ we get
$$
\widehat{\Omega} (\pi,\psi) = i \bigg[ \psi\psi^* \left(%
\begin{array}{cc}
  1 & 0 \\
  0 & 1 \\
\end{array}%
\right) + {\hbar\over 2} \left(%
\begin{array}{cc}
  1 & 0 \\
  0 & -1 \\
\end{array}%
\right) + \psi^* \sqrt{\hbar} \left(%
\begin{array}{cc}
  0 & 1 \\
  0 & 0 \\
\end{array}%
\right) - \psi \sqrt{\hbar} \left(%
\begin{array}{cc}
  0 & 0 \\
  1 & 0 \\
\end{array}%
\right) \bigg]
$$
\begin{equation}
 = i \bigg( \psi \psi^* - \psi \widehat{\psi}^* +
\psi^* \widehat{\psi} - \widehat{\psi}^* \widehat{\psi} +
{\hbar\over 2} \bigg).
\label{SW}
\end{equation}
$\big($Remember that $\varepsilon_{_{_{\left(%
\begin{array}{c}
  0 \\
  1 \\
\end{array}%
\right)}}} = \varepsilon_{_{_{\left(%
\begin{array}{cc}
  0 & 1 \\
\end{array}%
\right)}}} = 1!$ $\big).$\\
Then by Eqs. (\ref{state}) and (\ref{SW}) one quickly finds that
\begin{equation}
{\rm tr}\bigg\{ \widehat{\Omega} (\pi,\psi) \bigg\} = {1\over
i\hbar} \int d\psi' \langle \psi' | \widehat{\Omega} (\pi,\psi)
|\psi' \rangle=1
\end{equation}
as should be. From (\ref{SW}) we obtain that $\widehat{\Omega}$
satisfies the relation
\begin{equation}
\widehat{\Omega}^* (\pi,\psi) = - \widehat{\Omega} (\pi,\psi).
\end{equation}
One can also check that
\begin{equation}
{\rm tr}\bigg\{ \widehat{\Omega} (\pi',\psi') \widehat{\Omega}
(\pi'',\psi'') \bigg\} = \delta(\psi'-\psi'',\pi' - \pi'').
\end{equation}

\vskip 1 truecm \noindent {\it The Moyal $\star$-Product}

The Moyal $\star$-product is given by
$$
\star = \exp \bigg\{  {i\hbar\over 2} \bigg( {{\buildrel
\leftarrow \over \partial} \over \partial\psi}
{{\buildrel\rightarrow \over
\partial} \over \partial\pi} + {{\buildrel \leftarrow \over
\partial} \over \partial\pi} {{\buildrel\rightarrow \over
\partial} \over \partial\psi}
\bigg)\bigg\}= \exp \bigg\{  {\hbar\over 2} \bigg( {{\buildrel
\leftarrow \over \partial} \over \partial\psi}
{{\buildrel\rightarrow \over
\partial} \over \partial\psi^*} + {{\buildrel \leftarrow \over
\partial} \over \partial\psi^*} {{\buildrel\rightarrow \over
\partial} \over \partial\psi}
\bigg)\bigg\}
$$
\begin{equation}
= \exp \bigg\{  {\hbar\over 2} \bigg( {{\buildrel \leftarrow \over
\partial} \over \partial Q} {{\buildrel\rightarrow \over
\partial} \over \partial Q} + {{\buildrel \leftarrow \over
\partial} \over \partial P} {{\buildrel\rightarrow \over
\partial} \over \partial P}
\bigg)\bigg\}, \label{moyalstar}
\end{equation}
where $Q$ and $P$ are the oscillator variables (\ref{sw3}).

\vskip 1 truecm \noindent {\it The Wigner Function}

Consider the eigenvalue equation for the Hamiltonian $H$ given
by (\ref{ham_omega})
\begin{equation}
H * \rho_{_W} = E \rho_{_W},
\label{shcrodinger}
\end{equation}
where $\rho_{_W}$ stands for the Wigner function. We need also
that $\rho_{_W}$ be a {\it real} Grassmann superfunction such that
\begin{equation}
\rho_{_W}^* = \rho_{_W}.
\end{equation}
Therefore it expands as
\begin{equation}
\rho_{_W} = A_0 + A_1 \psi^* \psi, \label{wigner}
\end{equation}
where $A_0,$ and  $A_1$ are real numbers. Substituting
(\ref{ham_omega}), (\ref{moyalstar}) and (\ref{wigner}) into
(\ref{shcrodinger}) we get two linear equations for $A_0$ and
$A_1$
\begin{equation}
\begin{array}{l}
EA_0 - {\hbar^2\over 4}\omega A_1 =0,\\
-\omega A_0 + EA_1 =0.
\end{array}
\label{eqnlin}
\end{equation}
As $|A_0| + |A_1| \neq 0$ one finds
\begin{equation}
\det\left(%
\begin{array}{cc}
  E & -{\hbar^2\over 4}\omega \\
  -\omega & E \\
\end{array}%
\right) =0, \label{energy}
\end{equation}
with the solutions
\begin{equation}
E=\mp {\hbar\omega\over 2}.
\end{equation}
Then (\ref{eqnlin}) and (\ref{energy}) lead to the following
energy eigenvalues
\begin{equation}
E^{^{(-)}} = -{\hbar\omega\over 2},  \ \ \ \ \ E^{^{(+)}} =
{\hbar\omega\over 2},
\label{nivels}
\end{equation}
with corresponding eigenfunctions
\begin{equation}
\rho_{_W}^{^{(-)}} = A_0(1-{2\over\hbar} \psi^* \psi), \ \ \ \ \
\rho_{_W}^{^{(+)}} = A_0(1+{2\over\hbar} \psi^* \psi).
\label{wignersusy}
\end{equation}

As is known the density operator is given by
\begin{equation}
\widehat{\rho} = \int \rho_{_W} (\pi,\psi) \widehat{\Omega}
(\pi,\psi) d\pi d\psi.
\end{equation}
Then (with the usual trace ${\rm Tr}$) we have
\begin{equation}
\begin{array}{l}
{\rm Tr} \{\widehat{\rho}\} = \int \rho_{_W} (\pi,\psi)\bigg[ {\rm
Tr}
\bigg\{\widehat{\Omega} (\pi,\psi) \bigg\}\bigg] d\pi d\psi\\
\ \ \ \ = \int  \rho_{_W} (\pi,\psi) 2 \delta(\psi) \delta({\pi}) d\pi d\psi\\
 \ \ \ = 2 \rho_{_W} (0,0).
\end{array}
\end{equation}
But since Tr$\{\widehat{\rho}\} = 1,$ this implies that $A_0=
\rho_{_W} (0,0) ={1\over 2}.$ Substituting this value of $A_0$
into (\ref{wignersusy}) we get
\begin{equation}
\rho_{_W}^{^{(-)}} = {1\over 2} (1-{2\over\hbar} \psi^* \psi) =
  {1\over 2} \exp \bigg\{ -{2\over\hbar} \psi^*\psi \bigg\},  \ \
  \ \ \
\ \  \rho_{_W}^{^{(+)}} = {1\over 2} (1+{2\over\hbar} \psi^* \psi)
={1\over 2} \exp \bigg\{ {2\over\hbar} \psi^*\psi \bigg\}.
\label{nivelsofenergy}
\end{equation}
Finally, from Eqs. (\ref{SW}) and (\ref{nivelsofenergy}) after
some algebra we find
$$
\widehat{\rho}_{_W}^{^{(-)}} = \int \rho_{_W}^{^{(-)}} (\pi,\psi)
\widehat{\Omega} (\pi,\psi) d\pi d\psi
$$
$$
= {1\over 2} \int \bigg( \psi\psi^* - {2\over\hbar} \cdot
{\hbar\over 2} \psi^*\psi  + {2\over\hbar} \psi^* \psi
\widehat{\psi}^* \widehat{\psi} \bigg) d\psi^* d\psi
$$
\begin{equation}
= 1- {1\over\hbar} \widehat{\psi}^* \widehat{\psi} = \left(%
\begin{array}{cc}
  1 & 0 \\
  0 & 0 \\
\end{array}%
\right) = \left(%
\begin{array}{c}
  1 \\
  0 \\
\end{array}%
\right)\otimes \left(%
\begin{array}{cc}
  1 & 0 \\
\end{array}%
\right)= | 0 \rangle \langle 0|.
\end{equation}
Analogously for $\rho_{_W}^{^{(+)}} (\pi,\psi)$ we get
\begin{equation}
\widehat{\rho}^{(+)} = {1\over\hbar} \widehat{\psi}^*
\widehat{\psi} = \left(%
\begin{array}{cc}
  0 & 0 \\
  0 & 1 \\
\end{array}%
\right) = \left(%
\begin{array}{c}
  0 \\
  1 \\
\end{array}%
\right) \otimes \left(%
\begin{array}{cc}
  0 & 1  \\
\end{array}%
\right)\\  = | 1 \rangle \langle 1|.
\end{equation}

\vskip 1 truecm \noindent {\it Normal Ordering}

The Weyl correspondence of the hamiltonian (\ref{hamiltonian}) $H =
\omega \psi^* \psi= i \omega QP$ leads to the operator

$$
\widehat{H} = \int H(\pi,\psi) \widehat{\Omega} (\pi,\psi) d\pi
d\psi = i\omega \widehat{Q}\widehat{P}
$$
$$
= {1\over 2} \omega \bigg( \widehat{\psi}^* \widehat{\psi} -
\widehat{\psi} \widehat{\psi}^* \bigg) = \omega \bigg(
\widehat{\psi}^* \widehat{\psi} - {\hbar\over 2} \bigg)
$$
\begin{equation}
= {\hbar\omega\over 2} \left(%
\begin{array}{cc}
  -1 & 0 \\
  0 & 1 \\
\end{array}%
\right). \label{HAM}
\end{equation}
It is evident that
\begin{equation}
\widehat{H} |0\rangle = -{1\over 2} \hbar\omega |0\rangle,  \ \ \
\ \ \ \widehat{H} |1\rangle = {1\over 2} \hbar\omega |1\rangle.
\end{equation}
Then we can compute the normal ordered hamiltonian to be
$$
\widehat{H}_{\cal N} \equiv \ :\widehat{H}: \  = \int \big(
\widehat{\cal N} H \big) (\pi,\psi) \widehat{\Omega} (\pi,\psi)
d\pi d\psi
$$
$$
= \int \bigg[ \exp \bigg\{ {\hbar\over 2} {{\buildrel \rightarrow
\over \partial}^2 \over \partial\psi
\partial\psi^*} \bigg\} \omega \psi^*\psi \bigg]
\widehat{\Omega}(\pi,\psi) d\pi d\psi
$$
$$
= \int \bigg( \omega \psi^* \psi + {\hbar\omega\over 2} \bigg)
\widehat{\Omega} (\pi,\psi) d\pi d\psi
$$
$$
=i \omega \widehat{Q} \widehat{P} + {\hbar\omega \over 2} = \omega
\big( \widehat{\psi}^* \widehat{\psi} - {\hbar\over 2} \big) +
{\hbar\omega\over 2} = \omega \widehat{\psi}^* \widehat{\psi}
$$
\begin{equation}
= \hbar\omega \left(%
\begin{array}{cc}
  0 & 0 \\
  0 & 1 \\
\end{array}%
\right).
\end{equation}
Consequently, the eigenvalue equation
\begin{equation}
\widehat{H}_{\cal N} \widehat{\rho}' = E' \widehat{\rho}'
\end{equation}
is equivalent to
\begin{equation}
\big( \widehat{\cal N} H \big) \star \rho' = E' \rho'.
\label{schrnormal}
\end{equation}
Write $\rho' = A'_0 +
A'_1 \psi^* \psi$. Then the equation (\ref{schrnormal})
leads to the system of equations (compare with
(\ref{shcrodinger})-(\ref{wignersusy})):

\begin{equation}
\begin{array}{c}
\bigg( E'-{\hbar\omega\over 2} \bigg) A_0' -
{\hbar^2\omega\over 4} A'_1=0,\\
- \omega A'_0 + \bigg( E'-{\hbar\omega\over 2} \bigg) A'_1 = 0.
\end{array}
\end{equation}
Or equivalently

\begin{equation}
\det \left(%
\begin{array}{cc}
  E' - {\hbar\omega\over 2} & -{\hbar^2\omega\over 4} \\
  -\omega & E' - {\hbar\omega\over 2} \\
\end{array}%
\right) =0.
\end{equation}
Thus the eigenvalues of $H_{\cal N}$ read
\begin{equation}
E'^{^{(0)}}=0,  \ \ \ \ \ \ E'^{^{(+)}}= \hbar\omega.
\end{equation}
The corresponding Wigner functions are
\begin{equation}
\rho'^{^{(0)}}= {1\over 2} \bigg( 1-{2\over\hbar} \psi^*\psi
\bigg), \ \ \ \ \ \ \ \ \ \ \ \ \ \rho'^{^{(+)}}= {1\over 2}
\bigg( 1 + {2\over\hbar} \psi^*\psi \bigg).
\end{equation}
Finally, it is interesting to deal with the \textit{normal star}
product which is defined by
\begin{equation}
f \star_{_{(\mathcal{N})}} g = \widehat{\mathcal{N}}^{-1} \big(
\widehat{\mathcal{N}}f \star \widehat{\mathcal{N}} g \big).
\label{normalproduct}
\end{equation}
Consider the following eigenvalue equation
\begin{equation}
H \star_{_{(\mathcal{N})}} \rho'' = E'' \rho''.
\end{equation}
According to (\ref{normalproduct}) one has
\begin{equation}
\widehat{\mathcal{N}}^{-1} \big( \widehat{\mathcal{N}} H \star
\widehat{\mathcal{N}} \rho'' \big) = E'' \rho''.
\end{equation}
Hence $ \widehat{\mathcal{N}} H \star \widehat{\mathcal{N}} \rho''
= E'' \widehat{\mathcal{N}} \rho'' $ implies that
\begin{equation}
H_{\cal N} \star  \widehat{\mathcal{N}} \rho'' =
 E'' \widehat{\mathcal{N}} \rho''.
\end{equation}
Comparing with (\ref{schrnormal}) one quickly finds that $ E'' =
E'$ and $\rho'' \sim \widehat{\mathcal{N}}^{-1} \rho'.$ Consequently we
find that the eigenvalues
$$
E''^{^{(0)}} = 0, \ \ \ \ \ \ \ \  \ \ \ \ E''^{^{(+)}} =
\hbar\omega,
$$
correspond to the Wigner functions
\begin{equation}
\rho''^{^{(0)}} \sim \exp \bigg\{ - {\hbar\over 2} {{\buildrel
\rightarrow \over \partial}^2 \over
\partial\psi \partial \psi^*} \bigg\} \rho'^{^{(0)}} =
1-{1\over\hbar}\psi^* \psi, \ \ \ \ \ \rho''^{^{(+)}} \sim 1 +
{1\over\hbar}\psi^* \psi,
\end{equation}
respectively.
\subsection{Supersymmetric Weyl-Wigner-Moyal Formalism: The Susy Harmonic Oscillator}

In this subsection we construct a formalism for supersymmetric
quantum mechanics. To be more precise we construct all the
ingredients of the WWM formalism for a system defined on the
supersymmetric phase space of $2n \times 2N$ degres of freedom (or
super-phase-space) $\Gamma_{\cal S} =
\{(\vec{p},\vec{x},\vec{\pi},\vec{\theta})\}= \Gamma_B^{2n} \times
\Gamma_F^{2N}$; here $\Gamma_{B}^{2n}=\{(\vec{p},\vec{x})\}$ such
that $[\widehat{x}_j,\widehat{p}_k]_- = i\hbar\delta_{jk}$ and
$\Gamma_{F}^{2N}=\{(\vec{\pi},\vec{\theta})\}$ with
${[\widehat{\theta}_j,\widehat{\pi}_k]}_+ = i\hbar\delta_{jk}$. We
assume that $(\vec{p},\vec{x})$ are real bosonic variables and
$(\vec{\pi},\vec{\theta})$ are complex fermionic (Grassmann)
variables. We will find the Stratonovich-Weyl quantizer, the Moyal
$\star$-product and the Wigner function for the supersymmetric
systems. In particular one finds that the supersymmetric Moyal
$\star$-product is the tensor product of the corresponding Moyal
$\star$-products for the bosonic and fermionic systems. Then the
supersymmetric WWM machinery will be applied to the bosonic and
fermionic harmonic oscillators with $n=1$ and $N=1$. Let
$|{x},{\theta} \rangle= |{x}\rangle \otimes |{\theta} \rangle \in
{\cal H} $ be a coherent state of a quantum system in the
respective super Hilbert space ${\cal H} = {\cal H}_B \otimes
{\cal H}_F $. Define the normalized vacuum state $|0_B,0_F\rangle=
|0_B\rangle \otimes |0_F\rangle$ as
\begin{equation}
\widehat{\theta_j}|0_B,0_F\rangle = 0,  \ \ \ \ \ \ \ \ \ \ \ \ \
\widehat{x_j}|0_B,0_F\rangle=0.
\end{equation}
Analogously to (\ref{seis}) we have

\begin{eqnarray}
|x,\theta\rangle &=&\exp \bigg\{-{i\over
\hbar}\bigg[\sum_{\alpha=1}^n\widehat{p}_\alpha x_\alpha +
\sum_{j=1}^N
\widehat{\pi}_j \theta_j \bigg]\bigg\}|0_B,0_F\rangle\nonumber\\
&=&\exp\bigg\{-{i\over \hbar}\sum_{\alpha=1}^n\widehat{p}_\alpha
x_\alpha \bigg\}|0_B,{\theta}\rangle \nonumber\\
&=& \exp\bigg\{-{i\over \hbar}\sum_{j=1}^N\widehat{\pi}_j
\theta_j\bigg\}|{x},0_F\rangle.
\end{eqnarray}
This state satisfies the following eigenvalue equations
\begin{eqnarray}
\widehat{x}_k|{x},{\theta}\rangle = x_k|{x},{\theta}\rangle,\nonumber\\
\widehat{\theta}_k|{x},{\theta}\rangle =
\theta_k|{x},{\theta}\rangle.
\end{eqnarray}
Hence, for any superfunction
$\widehat{f}=f(\widehat{x},\widehat{\theta})$ we have
\begin{equation}
\widehat{f}|{x},{\theta}\rangle =
f(\widehat{x},\widehat{\theta})|{x},{\theta}\rangle
=f(x,\theta)|{x},{\theta}\rangle.
\end{equation}
Another useful result is the following (see (\ref{prop}))
\begin{eqnarray}
\exp\bigg\{-{i\over\hbar}\bigg[\sum_{\alpha=1}^n\widehat{p}_{\alpha}y_\alpha
+ \sum_{j=1}^N\widehat{\pi}_j\eta_j \bigg]\bigg\}
|{x},{\theta}\rangle = |{x}+ {y} ,{\theta} + {\eta} \rangle.
\end{eqnarray}
One can construct the corresponding bras $\langle {x},{\theta}|$ such that
\begin{eqnarray}
\langle x_\alpha,\theta_j | x_\beta,\theta_k \rangle &=& \langle x_\alpha| x_\beta \rangle \langle\theta_j |\theta_k \rangle\nonumber\\
&=&  \delta (x_\alpha-x_\beta) \delta(\theta_k-\theta_j).
\end{eqnarray}
The completeness relation reads
\begin{equation}
\int |{x},{\theta} \rangle {\cal D}x {\cal D}\theta \langle
{x},{\theta} | = 1.
\end{equation}
Note that the parity of the superstates is given by
\begin{equation}
\begin{array}{lcl}
\varepsilon_{|x\rangle}= \varepsilon_{|\theta\rangle} = 0 &
\Rightarrow
& \varepsilon_{|x,\theta\rangle} = 0,\\
\varepsilon_{\langle x|}= 0, \ \  \varepsilon_{\langle\theta|} = \left\{%
\begin{array}{ll}
    0 {\rm\ for\ }N{\rm\ even} \\
     1 {\rm\ for\ }N{\rm\ odd}
\end{array}%
\right.     & \Rightarrow & \varepsilon_{\langle x,\theta|}  = \left\{%
\begin{array}{ll}
    0 {\rm\ for\ }N{\rm\ even} \\
     1 {\rm\ for\ }N{\rm\ odd}
\end{array}%
\right..
\end{array}
\end{equation}

\vskip 1truecm \noindent {\it Stratonovich-Weyl Quantizer}

In what follows we restrict ourselves to the simplest case of
$n=1$ and $N=1$. The generalization to any $n$ and any $N$ is
obvious. Let $f=f(p,x,\pi,\theta)$ be a function on the
super-phase-space $\Gamma_{\cal S}$; its Fourier transform is
defined by
\begin{equation}
\widetilde{f} (\omega,\kappa,\lambda,\mu) := \int
f(p,x,\pi,\theta) \exp \{ -i (p \omega + x \kappa + \pi\lambda +
\theta \mu) \} dpdxd\pi d\theta.
\end{equation}
Inserting this into the Weyl quantization rule, the
corresponding operator $\widehat{f}$ is given by
\begin{equation}
\widehat{f} :=(2 \pi)^{-2} \int \widetilde{f} (\omega,\kappa,\lambda,\mu) \exp
\{ i (\widehat{p} \omega + \widehat{x} \kappa +
\widehat{\pi}\lambda + \widehat{\theta} \mu) \} d\omega d\kappa
d\lambda d\mu.
\end{equation}
After straightforward computations we obtain
\begin{equation}
\widehat{f} := (2 \pi \hbar)^{-1} \int f(p,x,\pi,\theta) \widehat{\Omega}
(p,x,\pi,\theta) dpdxd\pi d\theta,
\end{equation}
where $\widehat{\Omega} (p,x,\pi,\theta)$ is the supersymmetric
Stratonovich-Weyl quantizer

\begin{equation}
\widehat{\Omega} (p,x,\pi,\theta) = (2 \pi)^{-1} \hbar \int \exp \bigg\{ i
\bigg[(\widehat{p}-p)\omega + (\widehat{x}-x)\kappa +
(\widehat{\pi} - \pi) \lambda + (\widehat{\theta}- \theta) \mu
\bigg] \bigg\} d\omega d\kappa d\lambda d\mu.
\end{equation}
From its structure it is obvious that this is an even operator and
it can be rewritten in the form
\begin{equation}
\begin{array}{lll}
\widehat{\Omega} (p,x,\pi,\theta) &=& (2 \pi)^{-1} \hbar \int \exp \bigg\{ i
\bigg[(\widehat{p}-p)\omega + (\widehat{x}-x)\kappa\bigg]\bigg\}
 dp dx \otimes \int \exp \bigg\{ i \bigg[ (\widehat{\pi} - \pi) \lambda +
(\widehat{\theta}- \theta) \mu \bigg] \bigg\} d\pi d \theta\\
&=& \widehat{\Omega}_B (p,x) \otimes \widehat{\Omega}_F
(\pi,\theta).
\end{array}
\end{equation}
Another useful form of the supersymmetric Stratonovich-Weyl quantizer is
\begin{equation}
\begin{array}{lll}
\widehat{\Omega} (p,x,\pi,\theta) &=& {i\hbar}\int d\omega
d\lambda \exp\{ -{2i\over \hbar}\big( p \omega + \pi\lambda \big)
\}
|x-\omega,\theta-\lambda \rangle \langle x+\omega,\theta +\lambda|\\
&=&{i\hbar}\int d\omega d\lambda \exp\{ -{2i\over \hbar}\big(
p\omega + \pi\lambda \big) \} \big(|x-\omega\rangle \otimes
|\theta-\lambda \rangle \big) \big(\langle x+\omega| \otimes
\langle \theta
+ \lambda|\big)\\
&=&{i\hbar}\int d\omega d\lambda \exp\{ -{2i\over \hbar}\big(
p\omega + \pi\lambda \big) \} |x-\omega\rangle \langle x+\omega|
\otimes
|\theta-\lambda \rangle \langle \theta + \lambda|\\
&=& \widehat{\Omega}_B (p,x) \otimes \widehat{\Omega}_F
(\pi,\theta).
\end{array}
\end{equation}
The inverse Weyl correspondence is given by

\begin{equation}
f(p,x,\pi,\theta) = tr \{ \widehat{\Omega} (p,x,\pi,\theta)
\widehat{f} \},
\end{equation}
where the ``trace'' $tr$ is defined by
\begin{equation}
tr \{ \cdot \} := (i\hbar)^{-1} \int dxd\theta \langle x,\theta|
\cdot |x,\theta \rangle.
\end{equation}
The Wigner function corresponding to the density $\widehat{\rho}=|\Psi\rangle
\langle \Psi^{*}|$ associated with the superstate $|\Psi\rangle$
\begin{equation}
\begin{array}{lll}
\rho(p,x,\pi,\theta) &=& (2 \pi \hbar)^{-1} tr \{ \widehat{\Omega} (p,x,\pi,\theta)
\widehat{\rho} \}\\
&=& (2 \pi \hbar)^{-1} (i\hbar)^{-1} \int dx'd\theta' \langle x',\theta'|
\widehat{\Omega}
(p,x,\pi,\theta) \widehat{\rho}  |x',\theta' \rangle\\
&=& (2 \pi \hbar)^{-1} \int dx'd\theta' d\omega d\lambda \langle x',\theta'| \exp\{
-{2i\over \hbar}\big( p\omega + \pi\lambda \big) \}
|x-\omega,\theta-\lambda
\rangle \langle x+\omega,\theta +\lambda|\Psi\rangle \langle \Psi^{*}|x',\theta' \rangle\\
&=& (2 \pi \hbar)^{-1} \int d\omega d\lambda  \exp\{ -{2i\over \hbar}\big( p\omega +
\pi\lambda
\big) \} \Psi(x+\omega,\theta+\lambda) \Psi^{\dag}(x-\omega,\theta-\lambda) \\
\end{array}
\end{equation}
where $\Psi(x,\theta) = \langle x, \theta | \Psi \rangle$ and
$\Psi^{\dag}(x,\theta) = \langle\Psi^{*}|x,\theta\rangle$.

In the case of a supersymmetric pure state $|\Psi \rangle
= |\phi,\psi\rangle$ the Wigner function factorizes as follows
\begin{equation}
\begin{array}{lll}
\rho(p,x,\pi,\theta) &=& (2 \pi \hbar)^{-1} \int d\omega d\lambda  \exp\{ -{2i\over
\hbar}\big( p\omega + \pi\lambda \big) \}
\Psi(x+\omega,\theta+\lambda)
\Psi^{\dag}(x-\omega,\theta-\lambda) \\
&=& (2 \pi \hbar)^{-1}  \int d\omega d\lambda  \exp\{ -{2i\over \hbar}\big( p\omega +
\pi\lambda \big) \} \phi(x+\omega)\psi(\theta+\lambda)
\phi^{\dag}(x-\omega)
\psi^{\dag}(\theta-\lambda)\\
&=& \rho_B(p,x)\cdot\rho_F(\pi,\theta).
\end{array}
\end{equation}

\vskip 1truecm \noindent {\it Supersymmetric Moyal Product}

The Moyal $\star$-product for any two functions in the super-phase
space is defined by
$$
\star_{\cal S} := \exp\bigg\{ {i\hbar\over 2}  {\buildrel
{\leftrightarrow} \over {\cal P}_{\cal S}} \bigg\}
$$
where ${\buildrel {\leftrightarrow} \over {\cal P}_{\cal S}}$ is
given by the relation
\begin{equation}
\begin{array}{lll}
{\buildrel {\leftrightarrow} \over {\cal P}_{\cal S}}&=&
{\buildrel {\leftrightarrow} \over {\cal P}}_B + {\buildrel
{\leftrightarrow}
\over {\cal P}}_F\\
&=&  \bigg( {{\buildrel {\leftarrow} \over
\partial}\over
\partial x}{{\buildrel {\rightarrow} \over \partial}\over
\partial p} - {{\buildrel {\leftarrow} \over \partial} \over
\partial p}{{\buildrel {\rightarrow} \over \partial} \over
\partial x} \bigg) +  \bigg({\overleftarrow{\partial } \over
\partial\pi} {\overrightarrow{\partial} \over \partial \theta} +
{\overleftarrow{\partial} \over \partial \theta}
{\overrightarrow{\partial} \over
\partial\pi}\bigg). \label{poissonsusy}
\end{array}
\end{equation}
This is exactly the super-Poisson bracket operator used in Refs.
\cite{berezinbook,casa,marinov,sberezin,dewittbook,susyqmb,
bordemann,bordemannII,duetsch,zachosfermions,hirshfield,clifford}.

\vskip 1truecm \noindent {\it Normal Ordering}

Normal ordering corresponding to a supersymmetric system is the
composition of the normal ordering for bosons, and the normal
ordering for fermions i.e.,
\begin{eqnarray}
\widehat{\cal N}_S &=& \widehat{\cal N}_B\widehat{\cal N}_F=
\exp \left\{-{1 \over 2}\left(
{{\buildrel{\rightarrow}\over{\partial}}^2 \over \partial a
\partial a^*} -
{{\buildrel{\rightarrow}\over{\partial}}^2 \over \partial b
\partial b^*}\right)\right\},
\end{eqnarray}
where $a,a^*,b,b^*$ are the oscillator variables, which will be
defined later.

\vskip 1truecm \noindent {\it The System}

To begin with we define the lagrangian for a supersymmetric
harmonic oscillator of one variable and of a
frequency $\omega_B$ for the bosonic oscillator and $\omega_F$ for
the Fermi one
\begin{equation}
L= {1\over 2} p^2 - {1\over 2}\omega_B^2x^2 + i
\theta^*\dot{\theta} - \omega_F \theta^*\theta.
\end{equation}
Then the hamiltonian reads
\begin{equation}
H={1\over 2} p^2 + {1\over 2}\omega_B^2x^2 + \omega_F
\theta^*\theta.
\end{equation}
and the respective hamiltonian operator is
\begin{equation}
\widehat{H}={1\over 2} \widehat{p}^2 + {1\over
2}\omega_B^2\widehat{x}^2 + \frac{\omega_F}{2}( \widehat{\theta}^*
\widehat{\theta} - \widehat{\theta}
\widehat{\theta}^*)\label{Hsusy}.
\end{equation}
Here $\widehat{x},\widehat{p},\widehat{\theta},\widehat{\theta^*}$
obey the standard rules
$$
[\widehat{x},\widehat{p}]_- = i\hbar, \\ \ \ \ \ \
{[\widehat{x},\widehat{x}]_- }= {[\widehat{p},\widehat{p}]_-} = 0,
$$
$$
{[\widehat{\theta},\widehat{\theta}^*]}_+ = \hbar, \ \ \ \ \ \ \
{[\widehat{\theta},\widehat{\theta}]}_+ =
{[\widehat{\theta}^*,\widehat{\theta}^*]}_+ = 0.
$$
According to the standard procedure let us  define:
$$
\widehat{a}= {\sqrt{\omega_B\over 2\hbar}}\bigg(\widehat{x} +
{i\widehat{p}\over \omega_B}\bigg), \ \ \ \ \ \ \ \ \
\widehat{a}^*= {\sqrt{\omega_B\over 2\hbar}}\bigg(\widehat{x} -
{i\widehat{p}\over \omega_B}\bigg);
$$
$$
\widehat{b}={\widehat{\theta}\over \sqrt{\hbar}}, \ \ \ \ \ \ \ º
\ \ \widehat{b}^* = {-i\over \sqrt{\hbar}}\widehat{\pi}=
{\widehat{\theta}^*\over \sqrt{\hbar}}
$$
or
$$
\widehat{x}= {\sqrt{\hbar\over 2\omega_B}}\bigg(\widehat{a}^* +
\widehat{a}\bigg), \ \ \ \ \ \ \ \ \ \  \widehat{p}=
{i\sqrt{\hbar\omega_B\over 2}}\bigg(\widehat{a}^* - \widehat{a}
\bigg)
$$
$$
\widehat{\theta}=\sqrt{\hbar}\widehat{b}, \ \ \ \ \ \ \ \ \ \ \
\widehat{\theta}^* = -i \widehat{\pi}= \sqrt{\hbar}\widehat{b}^*;
$$
Then
$$
[\widehat{a},\widehat{a}^*]_- = 1, \ \ \ \ \ \ \
{[\widehat{a},\widehat{a}]_- }={[\widehat{a}^*,\widehat{a}^*]_-} =
0,
$$
$$
{[\widehat{b},\widehat{b}^*]}_+ = 1, \ \ \ \ \ \ \ \ \
{[\widehat{b},\widehat{b}]}_+ = {[\widehat{b}^*,\widehat{b}^*]}_+
= 0.
$$

With the use of these operators one gets
\begin{equation}
\widehat{H}={\hbar\omega\over 2}{[\widehat{a}^*,\widehat{a}]}_+ +
{\hbar\omega\over 2}{[\widehat{b}^*,\widehat{b}]_-},
\end{equation}
and finally
\begin{eqnarray}
\widehat{H}&=&\hbar\omega(\widehat{a}^*\widehat{a} + 1/2 +
\widehat{b}^*\widehat{b}-1/2),\nonumber\\
&=&\hbar\omega(\widehat{n}_B + \widehat{n}_F),
\end{eqnarray}
where due to the condition of unbroken supersymmetry (i.e. the
total energy of the vacuum vanishes) we put
$\omega_B=\omega_F=\omega$ and the normal ordering is no longer
necessary in this case; $\widehat{n}_B= \widehat{a}^*\widehat{a}$
and $\widehat{n}_F=\widehat{b}^*\widehat{b}$ are the number
operator for bosons and fermions, respectively.

On the other hand, in the spirit of supersymmetric quantum
mechanics, we can construct the supercharges as follows
\begin{equation}
\widehat{\cal Q}= \sqrt{2\omega}\ \widehat{a}^* \ \widehat{b},
\hspace{1.5cm} \widehat{\cal Q}^*= \sqrt{2\omega}\ \widehat{a}\
\widehat{b}^*,
\end{equation}
In terms of these supercharges

\begin{equation}
\widehat{H}= {\hbar \over 2}( \widehat{\cal Q} \widehat{\cal Q}^*
+ \widehat{\cal Q}^* \widehat{\cal Q}).
\end{equation}
The commutation relations read
\begin{eqnarray}
{[\widehat{\cal Q},\widehat{\cal Q}]}_+ &=&
{[\widehat{\cal Q}^*,\widehat{\cal Q}^*]}_+ =
0,\nonumber\\
{[\widehat{\cal Q},\widehat{H}]}_- &=& {[\widehat{\cal Q},\widehat{H}]}_- = 0.
\end{eqnarray}

The super-Weyl correspondence gives
$$
a= tr \{\widehat{\Omega}_{\cal S}\widehat{a}\}= {\sqrt{\omega\over
2\hbar}}\bigg({x} + {i{p}\over \omega}\bigg), \ \ \ \ \ \ \ \ \ \
a^* = tr \{ \widehat{\Omega}_{\cal S}\widehat{a}^*\} =
{\sqrt{\omega\over 2\hbar}}\bigg({x} - {i{p}\over \omega}\bigg),
$$
\begin{equation}
b= tr\{\widehat{\Omega}_{\cal S}\widehat{b}\}={{\theta}\over
\sqrt{\hbar}}, \ \ \ \ \ \ \ \ \ b^* = tr\{ \widehat{\Omega}_{\cal
S} \widehat{b}^*\} = -{i\over \sqrt{\hbar}}{\pi}= {{\theta}^*\over
\sqrt{\hbar}}.
\end{equation}
and
\begin{equation}
H= \hbar\omega \big(a^*\star_{\cal S} a + b^*\star_{\cal S} b
\big) = \hbar \omega \big(a^*a + b^*b \big).\label{hsusy}
\end{equation}
where $ \star_{\cal S}= \exp\{{i\hbar\over 2}
{\buildrel{\leftrightarrow}\over{\cal{P}_{\cal S}}} \}$ and
\begin{equation}
{\buildrel{\leftrightarrow}\over{\cal{P}_{\cal S}}}= -{i\over
\hbar} \bigg(- {{\buildrel {\leftarrow}\over \partial}\over
\partial a^*} {{\buildrel
{\rightarrow}\over \partial}\over
\partial a} + {{\buildrel {\leftarrow}\over \partial}\over
\partial a} {{\buildrel
{\rightarrow}\over \partial}\over
\partial a^*} + {{\buildrel
{\leftarrow}\over \partial}\over
\partial b^*}{{\buildrel
{\rightarrow}\over \partial}\over
\partial b}
 + {{\buildrel {\leftarrow}\over \partial}\over
\partial b}{{\buildrel
{\rightarrow}\over \partial}\over
\partial b^*}  \bigg)\label{pois}
\end{equation}
The hamiltonian splits into the bosonic and
fermionic parts
\begin{equation}
H=H_B + H_F.
 \label{Hsusy}
\end{equation}
For the supercharge operators the super-Weyl correspondence gives
\begin{eqnarray}
{\cal Q}&=& tr\{ \sqrt{2\omega}\ \widehat{a}^* \ \widehat{b} \} =
\sqrt{2\omega}\ a^*\star_{\cal S} b = \sqrt{2\omega}\ a^*b, \hspace{1.0cm}\nonumber\\
{\cal Q}^*&=& tr\{ \sqrt{2\omega}\ \widehat{a} \ \widehat{b}^* \}
= \sqrt{2\omega}\ a\star_{\cal S} b^* = \sqrt{2\omega}\ ab^*,
\end{eqnarray}
and then
$$
{\cal Q} \star_{\cal S} {\cal Q} = 0, \ \ \ \ \ \ \ {\cal
Q}^*\star_{\cal S}Q^* = 0,
$$
\begin{equation}
\hbar ({\cal Q}\star_{\cal S}{\cal Q}^* + {\cal Q}^* \star_{\cal
S}{\cal Q}) = 2\hbar\omega (a^*a+ b^*b)= 2H.
\label{charges}
\end{equation}
Finally the Wigner function of the ground state satisfies the equation
\begin{equation}
H\star_{\cal S} {\rho_W}_0 = E_0 {\rho_W}_0  =0. \label{hamilt}
\end{equation}
One can easily find that the solution of this equation is given by
\begin{equation}
{\rho_W}_0= {{\rho_W}_B}_0{{\rho_W}_F}_0\propto \exp\{-2(a^*a +
b^*b)\}.
\end{equation}
\vskip 2truecm
\section{Final Remarks}

In this paper we have studied the Weyl-Wigner-Moyal formalism for
fermionic systems with a finite number of degrees of freedom.

The relevant objects involved in the WWM-formalism as the
Stratonovich-Weyl quantizer, the Moyal $\star$-product, the Wigner
functions and the normal ordering have been found. Two examples have been
discussed in detail: the Fermi oscillator and the supersymmetric
quantum mechanics. The relation to the results of Ref.
\cite{hirshfield} is also explicitly given. In this respect, our
results are complementary to those of \cite{hirshfield}.

The extension of our present considerations to the fermionic systems
with an infinite number of degrees of freedom will be reported in a future
communication \cite{dirac}. In particular we are going to deal with
deformation quantization of fermionic fields coupled to
electromagnetic fields  \cite{campos},
\cite{qed}.

\vskip 2truecm
\centerline{\bf Acknowledgments}

It is a pleasure to thank G. Dito and D. Sternheimer for
stimulating comments and suggestions. This work was supported in
part by CONACyT M\'exico Grants 41993-F and 45713-F. I.G. wish to
thank Cinvestav, Unidad Monterrey for its hospitality where part
of this work was done. The research of I.G. is supported by a
CONACyT graduate fellowship.




\end{document}